\documentstyle[12pt]{article}

\input epsf
\input{tcilatex}

\begin{document}
\title{\bf Non Standard Extended Noncommutativity of Coordinates}
\author{\\A. BOULAHOUAL  and  M. B. SEDRA\footnote{Corresponding author: sedra@ictp.trieste.it}\\
\small{Abdus Salam International Centre For Theoretical Physics
ICTP, Trieste, Italy}}
 \maketitle \hoffset=-1cm
\textwidth=11,5cm \vspace*{1cm}

\begin{abstract}
We present in this short note an idea about a possible extension of the
standard noncommutative algebra of coordinates to the formal differential
operators framework. In this sense, we develop an analysis and derive an
extended noncommutative structure given by $[x_{a},x_{b}]_{\star }=i(\theta
+\chi )_{ab}$ where $\theta _{ab}$ is the standard noncommutativity
parameter and $\chi _{ab}(x)\equiv \chi _{ab}^{\mu }(x)\partial _{\mu }=%
\frac{1}{2}(x_{a}\theta ^{\mu }\,_{b}-x_{b}\theta ^{\mu }\,_{a})\partial
_{\mu }$ is an antisymmetric non-constant vector-field shown to play the
role of the extended deformation parameter. This idea was motivated by the
importance of noncommutative geometry framework, with nonconstant
deformation parameter, in the current subject of string theory and D-brane
physics.
\end{abstract}
{\bf Keywords}:Star product, differential operators,
Noncommutative algebra, string theory.
\hoffset=-1cm \textwidth=11,5cm \vspace*{1cm}

\hoffset=-1cm \textwidth=11,5cm \vspace*{1cm}

\hoffset=-1cm \textwidth=11,5cm

\vspace*{0.5cm}

\newpage

\section{Introduction}

Recently there has been a revival interest in the noncommutativity of
coordinates in string theory and D-brane physics[1-6]. This interest is
known to concern also noncommutative quantum mechanics and noncommutative
field theories [7, 8]. Before going into presenting the aim of our work, we
will try in what follows to expose some of the results actually known in
literature.\newline
The sharing property between all the above interesting areas of research is
that the corresponding space exhibits the following structure

\begin{equation}
\begin{array}{lcr}
\lbrack x_{i},x_{j}]_{*^{\prime }}=i\theta _{ij} &  &
\end{array}
\end{equation}
where $x_{i}$ are non-commuting coordinates which can describe also the
space-time coordinates operators and $\theta _{ij}$ is a constant
antisymmetric tensor. Quantum field theories living on this space are
necessarily noncommutative field theories. Their formulation is simply
obtained when the algebra (1) is realized in the space of fields (functions)
by means of the Moyal bracket according to which the usual product of
functions is replaced by the star-product as follows [9]

\begin{equation}
(f*g)(x) = f(x)e^{\frac{i}{2} \theta^{ab} \overleftarrow{\partial_a}%
\overrightarrow{\partial_b}} g(x) ,
\end{equation}
The link with string theory consist on the correspondence between the $%
\theta^{ij}$-constant parameter and the constant antisymmetric two-form
potential $B^{ij}$ on the brane as follows [1].

\begin{equation}
\theta^{ij} = (\frac{1}{B})^{ij},
\end{equation}
such that in the presence of this $B$-field, the end points of an open
string become noncommutative on the D-brane.\newline
In this letter, we try to go beyond the standard noncommutative algebra (1)
by presenting some computations leading to consider among other a
non-constant antisymmetric $\widehat\theta$-parameter satisfying an extended
noncommutative Heisenberg-type algebra given by
\begin{equation}
[x_{i}, x_{j}]_{\star}= i \widehat\theta_{ij}(x),
\end{equation}
where $\widehat\theta_{ij}(x)=(\theta +\chi)_{ij}$ and where $\chi_{ij}(x) =
\chi_{ij}^{\mu}(x)\partial_{\mu}=\frac{1}{2}(x_a \theta^{\mu}\,_{b} - x_b
\theta^{\mu} \, _{a})\partial_\mu$ describes a non-constant vector-field
deformation parameter. This is important since the obtained algebra (4) can
be reduced to the standard noncommutative algebra (1) once one forget about
the operatorial part of $\widehat\theta$ namely $\chi_{ij} =
\chi_{ij}^{\mu}\partial_{\mu}$. This construction is also interesting as it
may help to build a correspondence between the noncommutative geometry
framework, based on the algebra (4), and the string theory with a
non-constant B-field.

\section{Non standard noncommutative algebra.}

Consider the noncommutative space defined by the relation (1) originated
from the star product definition of two functions $f$ and $g$ of an algebra $%
{\cal A}$ that is given by
\begin{equation}
(f*^{\prime}g)(x) = f(x)e^{\frac{i}{2} \theta^{ab} \overleftarrow{\partial_a}
\overrightarrow{\partial_b}} g(x) ,
\end{equation}
where $\partial_a = \frac{\partial}{\partial {x^{a}}}$. We denote the
star-product in (5) by a prime for some reasons that we will explain later.
With this star product, one can define the Moyal bracket as follows
\begin{equation}
[f(x),g(x)]_{\ast^{\prime}} = f(x)*^{\prime}g(x) - g(x)*^{\prime}f(x).
\end{equation}
For functions $f$ and $g$ coinciding with the coordinates $x_{i}$ and $x_{j}$%
, we recover in a simple way (1). Actually, our idea starts from the
observation that the derivatives $\partial _{a} = \frac{\partial}{\partial
x^{a}}$ in the exponential (2) are differential operators which act in the
following way:
\begin{equation}
\partial _{a} : {\cal A} \rightarrow {\cal A} ,
\end{equation}
such that the prime derivative is given by
\begin{equation}
\begin{array}{lcl}
\partial_{a}f & = & f^{\prime}_{a} \\
\partial_{a}(fg) & = & f^{\prime}_{a}g+fg^{\prime}_{a} \\
\partial_{a}x_{i} & = & \delta_{ai}.
\end{array}
\end{equation}
Furthermore, for two given functions $f$ and $g$ of the algebra ${\cal A}$,
the term $f*^{\prime}g$ remains an element of ${\cal A}$. So, the prime
introduced in the definition of the $\ast$-product (5) is just to express
the prime character of the derivative $\partial_a$ as shown in (7-8).
\newline

Looking for a possible generalization of the above analysis to the formal
differential operators framework, we shall now introduce another kind of
star-product, denoted by $\star$ and associated to an operatorial action of
the derivative $\partial_a$. Before going into describing how does it works,
let us first introduce the set $\Sigma^{(p,q)}$, $p \geq 0$ [10]. This is
the algebra of local differential operators of arbitrary spins and positive
degrees. The upper indices $(p,q)$ carried by $\Sigma$ are the lowest and
the highest degrees. A particular example is given by $\Sigma^{(0,0)}$ which
is nothing but the algebra ${\cal A}$, the structure usually used in the
standard $\ast$-product computations. Furthermore, in terms of the spin
quantum number $\Delta = s$, the space $\Sigma^{(p,q)}$ is given by

\begin{equation}
\Sigma ^{(p,q)}=\oplus _{s\in N}\Sigma _{s}^{(p,q)}.
\end{equation}
Typical elements of (9) are given for $(p,q)=(0,k)$ by
\begin{equation}
\Sigma {}_{s}^{(0,k)}=\sum_{m=0}^{k}\chi _{s-m}(x){}D^{m}=\sum_{m=0}^{k}\chi
_{s-m}^{\mu _{1}...\mu _{m}}(x){\partial _{\mu _{1}}...\partial _{\mu _{m}}}.
\end{equation}
Next, we assume that the derivative $\partial _{a}$ acts on the function $f$
as an operator in the following way
\begin{equation}
\partial _{a}f(x)=f_{a}^{\prime }(x)+f(x)\partial _{a},
\end{equation}
a fact which means that our derivative should be defined as

\begin{equation}
\partial _{a}:\Sigma ^{(0,i)}\rightarrow \Sigma ^{(0,i+1)}.
\end{equation}
This way to define the derivative is induced from the extended $\star $%
-product operation defined as follows

\begin{equation}
(f\star g)(x)=f(x).g(x)+\frac{i}{2}\theta ^{ij}(f_{i}^{\prime }+f\partial
_{i})(g_{j}^{\prime }+g\partial _{j})+...
\end{equation}
In section 3 we summarize some non trivial relation satisfied by the product
$\star $.\newline
The major difference between the two star products $*$ and $\star $ is that,
for a given function $f$ of the algebra ${\cal A}$, the term $\partial _{a}f$
belongs on the first case to the algebra ${\cal A}$ while on the second case
it is an element of the space $\Sigma ^{(0,1)}$. In general $(\partial
_{1}...\partial _{n}f)$ is an element of $\Sigma ^{(0,n)}$ which is a
particular set of the space of local differential operators denoted by $%
{\cal O(A)}$ and which we can realize as

\begin{equation}
{\cal O(A)}=\oplus _{0\leq p\leq q}\Sigma ^{(p,q)}({\cal A}),
\end{equation}
Now, consider the $\star $-product definition for the coordinates $x_{a}$
and $x_{b}$, we obtain by using the above analysis
\begin{equation}
\begin{array}{lcl}
x_{a}\star x_{b} & = & \sum\limits_{\alpha =0}^{\infty }(x_{a}\star
x_{b})_{\alpha } \\
& = & x_{a}x_{b}+\sum\limits_{{\alpha }=1}^{\infty }\frac{1}{\alpha !}(\frac{%
i}{2})^{\alpha }\prod\limits_{i=1}^{\alpha }\theta ^{\mu _{i}\nu
_{i}}(\prod\limits_{j=1}^{\alpha }\partial _{\mu
_{j}}x_{a})(\prod\limits_{k=1}^{\alpha }\partial _{\nu _{k}}x_{b})
\end{array}
\end{equation}
and explicitly we have,
\begin{equation}
\begin{array}{lcl}
(x_{a}\star x_{b})_{\alpha }=\frac{1}{\alpha !}(\frac{i}{2})^{\alpha } & %
\Bigg\{ & x_{a}x_{b}\prod\limits_{i=1}^{\alpha }(\theta ^{\mu _{i}\nu
_{i}}\partial _{\mu _{i}}\partial _{\nu _{i}}) \\
& + & \sum\limits_{i=1}^{\alpha }\Big\{\theta _{ab}\prod\limits_{k\ne
i}(\theta ^{\mu _{k}\nu _{k}}\partial _{\mu _{k}}\partial _{\nu _{k}}) \\
& + & x_{a}{\theta ^{\mu _{i}}\,_{b}}\prod\limits_{j\ne i}(\theta ^{\mu
_{j}\nu _{j}}\partial _{\nu _{j}})\prod\limits_{k}\partial _{\mu _{k}} \\
& + & (x_{a}\theta ^{\nu _{i}}\,_{b}+x_{b}\theta ^{\nu
_{a}}\,_{a})\prod\limits_{j\ne i}(\theta ^{\mu _{j}\nu _{j}}\partial _{\mu
_{j}})\prod\limits_{k}\partial _{\nu _{k}} \\
& + & \sum\limits_{j\ne i}\theta ^{\nu _{i}}\,_{a}\theta ^{\nu
_{j}}\,_{b}\prod\limits_{(k\ne i,k\ne j)}(\theta ^{\mu _{k}\nu _{k}}\partial
_{\mu _{k}})\prod_{l}\partial _{\nu _{l}} \\
& + & \sum\limits_{j\ne i}\theta _{a}\,^{\nu _{i}}\theta ^{\mu
_{j}}\,_{b}(\prod\limits_{l\ne j}\partial _{\nu _{l}})(\prod\limits_{(k\ne
i,k\ne j)}\theta ^{\mu _{k}\nu _{k}})(\prod\limits_{m\ne i}\partial _{\mu
_{m}})\Big\}\Bigg\}
\end{array}
\end{equation}
This result is obtained by using the derived recurrence formula (30).\newline
Furthermore, using the antisymmetry property of $\theta $, we can easily
check that (16) can be more simplified. Concrete examples are given by the
first term $x_{a}x_{b}\prod\limits_{i=1}^{\alpha }(\theta ^{\mu _{i}\nu
_{i}}\partial _{\mu _{i}}\partial _{\nu _{i}})$ which vanishes. Also $\theta
_{ab}\prod\limits_{k\ne i}(\theta ^{\mu _{k}\nu _{k}}\partial _{\mu
_{k}}\partial _{\nu _{k}}),$ $x_{a}{\theta ^{\mu _{i}}\,_{b}}%
\prod\limits_{j\ne i}(\theta ^{\mu _{j}\nu _{j}}\partial _{\nu
_{j}})\prod\limits_{k}\partial _{\mu _{k}}$ as well as $(x_{a}\theta ^{\nu
_{i}}\,_{b}+x_{b}\theta ^{\nu _{a}}\,_{a})\prod\limits_{j\ne i}(\theta ^{\mu
_{j}\nu _{j}}\partial _{\mu _{j}})\prod\limits_{k}\partial _{\nu _{k}}$ are
terms which contribute only for the value $\alpha =1$.\newline

On the other hand, performing straightforward but lengthy computations, we
find the following noncommutative extended $\star $-algebra
\begin{equation}
\lbrack x_{a},x_{b}]_{\star }=i\widehat{\theta }_{ab}(x),
\end{equation}
where the only non-vanishing term among a long mathematical series is given
by
\begin{equation}
\widehat{\theta }_{ab}=\theta _{ab}+\frac{1}{2}(x_{a}\theta ^{\mu
}\,_{b}-x_{b}\theta ^{\mu }\,_{a})\partial _{\mu }.
\end{equation}
Later on, we will denote the vector-field appearing on the rhs of (18)
simply by
\begin{equation}
\chi _{ab}\equiv \chi _{ab}^{\mu }(x)\partial _{\mu }=\frac{1}{2}%
(x_{a}\theta ^{\mu }\,_{b}-x_{b}\theta ^{\mu }\,_{a})\partial _{\mu },
\end{equation}
In this way, $\chi $ is interpreted as a deformation parameter term such
that the algebra (17) becomes
\begin{equation}
\lbrack x_{a},x_{b}]_{\star }=i\Big(\theta +\chi \Big)_{ab}.
\end{equation}
Note by the way that the long series we obtained for the $\widehat{\theta }$
-parameter, before simplifying to (18), is given by
\begin{equation}
\widehat{\theta }_{ab}(x)=\sum\limits_{\alpha =1}^{\infty }\widehat{\theta }%
_{ab}^{\alpha },
\end{equation}
with
\begin{equation}
\widehat{\theta }_{ab}^{\alpha }=\frac{1}{\alpha !}\frac{i^{\alpha -1}}{%
2^{\alpha }}\sum_{i=1}^{\alpha }\{2\theta _{ab}\prod_{k\ne i}(\theta ^{\mu
_{k}\nu _{k}}\partial _{\mu _{k}}\partial _{\nu _{k}})+(x_{a}\theta ^{\mu
_{i}}\,_{b}-x_{b}\theta ^{\mu _{i}}\,_{a})\prod_{j\ne i}(\theta ^{\mu
_{j}\nu _{j}}\partial _{\nu _{j}})\prod_{k=1}^{\alpha }\partial _{\mu
_{k}}\}.
\end{equation}

\section{Some Useful Formulas}

{\bf 1} let $c$ and $c^{\prime }$ be constant numbers, we have
\begin{equation}
\begin{array}{lcl}
c\star c^{\prime }=c.c^{\prime } &  &
\end{array}
\end{equation}
{\bf 2} For each function $f(x)$ on the algebra $\Sigma ^{(0,0)}\equiv {\cal %
A}$, we can show by using explicit computations that
\begin{equation}
\begin{array}{lcl}
f(x)\star c=f.c+\sum\limits_{\alpha =1}^{\infty }\frac{c}{\alpha !}(\frac{i}{%
2})^{\alpha }\prod_{i=1}^{\alpha }\theta ^{\mu _{i}\nu _{i}}f_{\mu
_{1}...\mu _{\alpha }}^{(\alpha )}\prod_{j=1}^{\alpha }\partial _{\nu _{j}}
&  &
\end{array}
\end{equation}
where for example $f_{a_{1}}^{(1)}$ is the prime derivative with respect to $%
\partial _{a_{1}}$\newline
{\bf 3} We have also
\begin{equation}
\begin{array}{lcl}
c\star f(x)=c.f(x) &  &
\end{array}
\end{equation}
{\bf 4} Combining (24-25) we find for the particular case $f=x$
\begin{equation}
\begin{array}{lcl}
\lbrack x_{\mu },c]_{\star }=\frac{i}{2}c.\theta _{\mu }\,^{\nu }\partial
_{\nu } &  &
\end{array}
\end{equation}
{\bf 5} Obviously
\begin{equation}
\begin{array}{lcl}
\partial _{a}\star \partial _{b}=\partial _{a}.\partial _{b} &  &
\end{array}
\end{equation}
{\bf 6} Also we have
\begin{equation}
\begin{array}{lcl}
\partial _{a}\star f(x) & = & \partial _{a}.f(x) \\
& = & f_{a}^{\prime }(x)+f\partial _{a}
\end{array}
\end{equation}
{\bf 7} The general formula
\begin{equation}
\begin{array}{lcl}
\prod_{j=1}^{\alpha }\partial _{\nu _{j}}\star f(x)=\prod_{j=1}^{\alpha
}\partial _{\nu _{j}}.f(x) &  &
\end{array}
\end{equation}
{\bf 8 } Applying to the coordinates $x_{a}$
\begin{equation}
\partial _{\mu _{1}}...\partial _{\mu _{n}}x_{a}=x_{a}\partial _{\mu
_{1}}...\partial _{\mu _{n}}+\sum\limits_{i=1}^{n}\delta _{a\mu
_{i}}\partial _{\mu _{1}}...\check{\partial}_{\mu _{i}}...\partial _{\mu
_{n}},
\end{equation}
or equivalently
\begin{equation}
(\prod\limits_{i=1}^{n}\partial _{\mu
_{i}})x_{a}=x_{a}(\prod\limits_{i=1}^{n}\partial _{\mu
_{i}})+\sum\limits_{i=1}^{n}\delta _{a\mu _{i}}(\prod\limits_{j\ne
i}^{n}\partial _{\mu _{j}}).
\end{equation}

\section{Concluding Remarks}

Following this construction, some remarks are in order:\newline
\newline
\{{\bf 1\}} A first important remark concerning the obtained algebra (17),
is that it does not closes as a standard algebra. This property is easily
observed since the extended Moyal bracket of $x_{a}$ and $x_{b}$; which are
coordinates elements of $\Sigma _{-1}^{(0,0)}={}{\cal A}_{-1}^{(0,0)}$;
gives $\widehat{\theta }_{ab}(x)=\theta _{ab}+\frac{1}{2}(x_{a}\theta ^{\mu
}\,_{b}-x_{b}\theta ^{\mu }\,_{a})\partial _{\mu }$ which is an element of $%
\Sigma _{-2}^{(0,1)}$. However, if we forget about the vector field term $%
\chi _{ab}^{\mu }\partial _{\mu }$ in $\widehat{\theta }_{ab}(x)$, we
recover the standard noncommutative structure (1) which is a closed algebra.
We can conclude for this point that the fact to transit from prime star
product $*$ to the operatorial one $\star ,$ is equivalent to introduce
local vector fields contributions at the level of the deformation parameter $%
\widehat{\theta }_{ab}$ which therefore becomes coordinates dependent.

\{{\bf 2\}} Related to \{{1\}}, we can also check that the associativity
with respect to the operatorial $\star $-product operation is not satisfied.
As an example consider
\begin{equation}
\begin{array}{lcl}
{(f\star 1)\star 1} & = & {}D(x)\star 1, \\
{f\star (1\star 1)} & = & {}D(x)
\end{array}
\end{equation}
where the differential operator ${}D(x)$ is just the result of $f\star 1$.
Then we can easily check that ${}D(x)\ne {}D(x)\star 1$ as shown in the
formulas (24-25). This property of non associativity of the operatorial star
product exhibits a particular interest. In fact it makes us recall the non
associative algebra based on results about open string correlation functions
proposed in [4a] and which deal with D-branes in a background with
non-vanishing H. \newline
{\bf \{3\}} Concerning the mentioned properties \{1-2\}, the problem of
closure of the derived algebra (17) can be approached by using the analogy
with the non-linear Zamolodchikov $W_{3}$-algebra which exhibits a similar
property. Namely the non-closure of the algebra due to the presence of the
spin-4 term in the commutation relations of $W_{3}$ currents. For a review
see [11].\newline
{\bf \{4\}} The noncommutative extended parameter $\widehat{\theta }_{ab}=%
\Big( \chi +\theta \Big)_{ab}$ is not a constant object contrary to $\theta
_{ab}$ and thus the associated algebra (17) is not a trivial structure as it
corresponds to a noncommutative deformation of the standard algebra (1) by
the vector fields $\chi _{ab}$.\newline
{\bf \{5\}} Using the derived relation (26), we can easily show that the non
constant deformation parameter $\chi _{ab}$ is given by $\chi _{ab}\equiv i%
\big \{ x_{b}.(x_{a}\star 1)-x_{a}.(x_{b}\star 1)\big \}$.\newline
{\bf \{6\}} $\theta _{ab}$ as well as the antisymmetric tensor $\chi
_{ab}(x) $ are objects of conformal weights $\Delta =-2$, since $\Delta
(\partial _{\mu })=-\Delta (x)=1$.\newline
{\bf \{7\}} From the mathematical point of view, $\widehat{\theta }%
_{ab}^{(\alpha )}$ given in (21-22) are general objects which belong to the
subspaces
\begin{equation}
\Sigma _{-2}^{(2\alpha -2,2\alpha -1)},
\end{equation}
and $\widehat{\theta }_{ab}$ given in (18) is nothing but the first
contribution for $\alpha =1$ and consequently is an object of $\Sigma
_{-2}^{(0,1)}$.\newline
\newline
{\bf \{8\}} We easily obtain the standard noncommutative algebra (1) from
(17) just by considering the following quotient space
\begin{equation}
\Sigma _{-2}^{(0,1)}\Big/ \Sigma _{-2}^{(1,1)}
\end{equation}
which consist simply on forgetting about vector fields $\chi ^{\mu }\partial
_{\mu }$.\newline
\newline

\newpage {\bf References}

\begin{enumerate}
\item[{[1]}]  A. Connes, M.R. Douglas, A. Schwarz, Noncommutative geometry
and matrix theory: Compactification on tori, JHEP 02(1998) 003,
[hep-th/9711162] and references there in.\newline
N. Seiberg, E. Witten, String Theory and Noncommutative geometry, JHEP 09
(1999) 032, [hep-th/9908142] and references there in.

\item[{[2]}]  C.S.Chu, P.M.Ho, Noncommutative open string and D-brane, Nucl.
Phys.B550, 151 (1999), [hep-th/9812219]\newline
M.R. Douglas, C. Hull, D-branes and the noncommutative Torus, JHEP 02 (1998)
008, [hep-th/9711165]\newline
B. Morariu, B. Zumino, in Relativity, Particle Physics and Cosmology, World
Scientific, Singapore, 1998, hep-th/9807198]\newline
W. Taylor, D-brane field theory on compact spaces, Phys. Lett. B394, 283
(1997), [hep-th/9611042].

\item[{[3]}]  Y.K.E. Cheung and M. Krogh, Noncommutative geometry from
0-branes in a background B field, Nucl. Phys. B 528 (1998)185,
[hep-th/9803031];\newline
F. Ardalan, H. Arfaei and M.M.Seikh-Jabbari, Noncommutative geometry from
strings and branes, JHEP 02(1999)016, [hep-th/9810072];\newline
M.M.Seikh-Jabbari, Open strings in a B field background as electric dipole,
Phys. Lett. B 455 (1999)129, [hep-th/9901080];\newline
M.M.Seikh-Jabbari, One Loop Renormalizability of Supersymmetric Yang-Mills
Theories on Noncommutative Two-Torus, JHEP 06 (1999)015, [hep-th/9903107];%
\newline
V. Schomerus, D-branes and deformation quantization, JHEP, 06(1999)030,
[hep-th/9903205];\newline
D. Bigatti and L. Susskind, Magnetic fields, branes and noncommutative
geometry, Phys.Rev. D62 (2000) 066004, [hep-th/9908056].

\item[{[4]}]  L. Cornalba and R. Schiappa, Non associative star product
deformations for D-brane world volume in curved backgrounds,
[hep-th/0101219] \newline
P.M. Ho, Y-T. Yeh, Noncommutative D-brane in nonconstant NS-NS B field
background, [hep-th/0005159], Phys. Rev. Lett. 85, 5523(2000)\newline
P.M. Ho, Making non associative algebra associative, [hep-th/0103024].

\item[{[5]}]  M. Aganagic, R. Gopakumar, S. Minwalla, A. Strominger,
Unstable Solitons in Noncommutative Gauge Theory, JHEP 0104 (2001)
001, [hep-th/0103256], \newline R. Gopakumar, J. Maldacena, S.
Minwalla, A. Strominger, S-Duality and Noncommutative Gauge
Theory, JHEP 0006 (2000) 036, [hep-th/0005048]\newline.

\item[{[6]}]  B. Jurco, P. Schupp, J. Wess, Nonabelian noncommutative gauge
theory via noncommutative extra dimensions [hep-th/0102129],\newline
B. Jurco, P. Schupp, J. Wess, Nonabelian noncommutative gauge fields and
Seiberg-Witten map, [hep-th/0012225],\newline
B. Jurco, P. Schupp, J. Wess, Noncommutative gauge theory for Poisson
manifolds, [hep-th/0005005],\newline

\item[{[7]}]  A. Micu and M.M.Seikh-Jabbari, Noncommutative $\Phi ^{4}$
Theory at Two Loops, JHEP 01(2001)025, [hep-th/0008057] and references there
in\newline
L. Bonora, M. Schnabl and A. Tomasiello, A note on consistent anomalies in
noncommutative YM theories, [hep-th/0002210].\newline
L. Bonora, M. Schnabl, M.M.Seikh-Jabbari, A. Tomasiello, Noncommutative $%
SO(n)$ and $Sp(n)$ Gauge theories, [hep-th/0006091].

\item[{[8]}]  M. Chaichian, M.M.Seikh-Jabbari, A. Tureanu, Hydrogen Atom
Spectrum and the Lamb Shift in Noncommutative QED, Phys. Rev. Lett.
86(2001)2716, [hep-th/0010175];\newline
M. Chaichian, A. Demichev, P. Presnajder, M.M.Seikh-Jabbari, A. Tureanu,
Quantum Theories on Noncommutative Spaces with Nontrivial Topology:
Aharonov-Bohm and Casimir Effects, [hep-th/0101209].

\item[{[9]}]  M. Kontsevitch, Deformation quantization of Poisson manifolds
I, [q-alg/9709040]\newline
D.B. Fairlie, Moyal Brackets, Star Products and the Generalized Wigner
Function, [hep-th/9806198]\newline
D.B. Fairlie, Moyal Brackets in M-Theory, Mod.Phys.Lett. A13 (1998) 263-274,
[hep-th/9707190],\newline
C. Zachos, A Survey of Star Product Geometry, [hep-th/0008010],\newline
C. Zachos, Geometrical Evaluation of Star Products, J.Math.Phys. 41 (2000)
5129-5134, [hep-th/9912238],\newline
C. Zachos, T. Curtright, Phase-space Quantization of Field Theory,\newline
Prog.Theor. Phys. Suppl. 135 (1999) 244-258, [hep-th/9903254],\newline
M.Bennai, M. Hssaini, B. Maroufi and M.B.Sedra, On the Fairlie's Moyal
formulation of M(atrix)- theory,Int. Journal of Modern Physics A V16(2001) ,
[hep-th/0007155].

\item[{[10]}]  E.H.Saidi, M.B.Sedra, On the Gelfand-Dickey algebra $GD(SLn)$
and the Wn-symmetry. I. The bosonic case, Jour. Math. Phys. V35N6(1994)3190%
\newline
M.B.Sedra, On the huge Lie superalgebra of pseudo-superdifferential
operators and super KP-hierarchies, Jour. Math. Phys. V37 (1996) 3483.

\item[{[11]}]  A. B. Zamolodchikov, Infinite additional symmetries in two
dimensional conformal quantum field theory, Teor. Math. Fiz. 65, 347 (1985)%
\newline
\end{enumerate}

\end{document}